\documentstyle[a4,cite,12pt,aps,epsfig]{revtex}

\makeatletter

\def\nnb{\nonumber}

\def\be{\begin{equation}}
\def\ee{\end{equation}}
\def\mn{\mu\nu}
\newcommand{\bea}{\begin{eqnarray}}
\newcommand{\eea}{\end{eqnarray}}
\makeatother

\begin{document}

\title{The renormalization of the
  effective gauge Lagrangian with
 spontaneous symmetry breaking: the U(1) case}

\author{
             {\bf Qi-Shu Yan}\footnote{
        E-mail Address: yanqs@mail.ihep.ac.cn} and {\bf Dong-Sheng Du}
\footnote{
        E-mail Address: duds@mail.ihep.ac.cn}  \\
	Theory division,
        Institute of high energy physics,
	Chinese academy of sciences, Beijing 100039,
	Peoples' Republic of China
}
\bigskip

\address{\hfill{}}

\maketitle

%\begin{center}
%Abstract
%\end{center}
\begin{abstract}
We study the renormalization of the nonlinear effective
$U(1)$ Lagrangian up to $O(p^4)$ with
spontaneous symmetry breaking.
The problems of the quartic divergences
and of the truncation of infinite divergence
tower are addressed. The renormalization group
equations of the effective Lagrangian are derived,
which make it possible to get the leading logarithm
corrections of the heavy degree of freedom, even
if the effective and full theories are matched at tree
level. The method we use in the $U(1)$ case can easily be extended to
study the non-Abelian effective theories and the electroweak chiral
Lagrangian.
\end{abstract}
\pacs{}

%\newpage
%\pagestyle{1}

To understand the nature, the effective field theory method
is a tool universal, practical, and powerful \cite{georgi, pich}.
For example, the Fermi weak interaction theory works quite well
at the energy scale below $m_W$ even before the standard model is
established. And the effective Hamiltonian method is
widely used in the B physics enterprise \cite{buras}.
Although the predictivity of a general effective theory
is restrained due to the fact that there are
infinite permitted operators in its Lagrangian,
at the region with energy lower enough than the ultraviolet cutoff,
these operators can be well organized in terms of
their importance to the low energy dynamics ( i.e. their dimension and the
strength of their couplings ). For example, among the
three groups of effective operators \cite{wilson}---the relevant,
marginal, and irrelevant ones---only the first two
groups dominate the dynamics of the low energy QED and QCD.
And in the Fermi theory and B physics theory, only operators equal
and below dimension 6 are important.

As one of important application of the effective field theory method,
the effective gauge theories with spontaneous symmetry breaking
play a very special role to describe the microscopic world, for example, the
QCD chiral Lagrangian \cite{gasser}, which describes the interactions among
hadrons, and the electroweak chiral Lagrangian \cite{appliq}, which describes
the interactions among massive vector bosons.
As we know, the renormalization group equations (RGEs) of an effective
theory are one of its basic ingredients
to describe the behavior of a given system, which
can effeciently sum up the logarithm corrections from the
quantum fluctuation of low energy degrees of freedom (DOF), eliminate or alleviate
the renormalization scale and scheme dependences, and improve
the perturbation method, especially when the radiative corrections
are significant, in B physics for instance. But for the effective
gauge theories with massive vector bosons ( and nonlinear interactions ),
it seems not easy to take into account the radiative corrections of low energy
quantum DOF.

This difficulty is more manifest when we represent the effective gauge
theories in their unitary gauge. The propagator of massive vector bosons
can be expressed as
\bea
i \Delta^{\mu\nu}&=&i \Delta_T^{\mu\nu}+i \Delta_L^{\mu\nu}\,,\\
\Delta_T^{\mu\nu}&=&\frac{1}{k^2-m_V^2} \left ( -g^{\mu\nu}+\frac{k^{\mu} k^{\nu}}{k^2} \right )\,,\\
\Delta_L^{\mu\nu}&=&\frac{1}{m_V^2} \frac{k^{\mu} k^{\nu}}{k^2}\,,
\eea
where $\Delta_T$ and $\Delta_L$ represent the transverse and
longitudinal parts, respectively.
The longitudinal part of the propagator can bring into
quartic divergences and lead to the well-known bad
ultraviolet behavior. Two direct consequences of this
fact are 1) that the quartic divergences will appear in
radiative corrections and 2) that low dimensions operators
can induce the infinite number of divergences of higher
dimension operators. In a renormalizable theory, the Higgs model for instance,
these two problems do not exist \cite{wood}. The quartic divergences produced by
the low engergy DOF just cancel exactly with
those produced by the Higgs scalar, and no extra
divergence structure will appear.

In this letter, we will show that these two problems can be
overcome in the framework of the effective
theory. And, indeed, we can
conduct the renormalization procedure order by order, as
we do in those renormalizable theories.
As an example to demonstrate the related conceptions and methods,
we extract the one-loop RGEs of the effective $U(1)$ gauge
theory up to the relevant and marginal ( $O(p^4)$ ) operators.

First, we briefly introduce the $U(1)$ Higgs model \cite{higgs} and
the effective Lagrangian ${\cal L}_{eff}$
with spontaneous symmetry breaking mechanism.
The partition functional of the renormalizable $U(1)$ Higgs model
can be expressed as
\bea
{\cal Z}=\int {\cal D}A_{\mu} {\cal D}\phi_1 {\cal D}\phi_2 \exp\left (i {{\cal S}[A,\phi_1,\phi_2]}\right )\,,
\eea
where the action ${\cal S}$ is determined by the following Lagrangian
\bea
{ \cal L} &=& -{1\over 4 e^2} F_{\mn}F^{\mn}
	 + (D\phi)^{\dagger}\cdot(D \phi)
	+\mu^2 \phi^{\dagger} \phi - {\lambda \over 4} (\phi^{\dagger} \phi)^2\,,
\eea
and the definition of quantities in the Lagrangian is given below
\bea
F_{\mn}&=&\partial_{\mu} A_{\nu} - \partial_{\nu} A_{\mu}\,,\\
D_{\mu} \phi&=&\partial_{\mu} \phi - i A_{\mu} \phi\,,\\
\phi&=&{1\over \sqrt{2}} (\phi_1 + i \phi_2)\,.
\eea
The mass square in the Higgs potential $\mu^2$ is positive,
as prescribed by hand.
And the vaccum expectation value of Higgs
field is nonzero and $\langle \phi \rangle = v/{\sqrt 2}$.
Then $U(1)$ symmetry is spontaneously broken, and
the vector bosons $A$ obtains their masses.

The non-linear realization of the Lagrangian with
spontaneous symmetry breaking is made by changing
the variable $\phi$
\bea
\phi={1\over \sqrt{2}} (v + \rho) U\,,\,\,
U=\exp\left ({i \xi \over v}\right )\,,\,\,
v=2 \sqrt{\mu^2 \over \lambda}\,,
\eea
then it reaches
\bea
{\cal L'} &=& -{1\over 4 e^2} F_{\mn}F^{\mn}
	 + {(v+\rho)^2\over 2} (D U)^{\dagger} \cdot (D U)
	+ {1\over 2} \partial \rho \cdot \partial \rho
	+{1\over 2} \mu^2 (v+\rho)^2 - {\lambda \over 16} (v+\rho)^4\,.
\eea
The field $U$ is the Goldstone boson as prescribed by the
Goldstone theorem, and $\rho$ is a
massive scalar field. And the change of variables
induces a determinant factor
in the functional integral
\bea
{\cal Z} = \int {\cal D}A_{\mu} {\cal D}\rho {\cal D}\xi \exp\left (i {\cal S}'[A,\rho,\xi] \right ) \det\left \{ \left  (1 + {1\over v} \rho\right  ) \delta(x-y) \right \}\,.
\eea
The determinant can be written in the exponential form (we can also represent
it into ghost terms),
and correspondingly the Lagrangian density is modified to
\bea
\label{hml}
{\cal L} \rightarrow {\cal L'} - i \delta(0) ln \left \{ 1 + {1\over v} \rho \right \}\,.
\label{uglag}
\eea
This determinant containing quartic divergences is
indispensible and crucial to cancel exactly the quartic
divergences caused by the longitudinal part
of vector bosons \cite{nish, wood}, and is important
in verifying the renormalizablity of the Higgs model in U-gauge.

In the low energy region, the effective dynamic DOFs
of the $U(1)$ gauge theory contain only the vector gauge bosons
and the Goldstone boson. The most general effective $U(1)$ gauge
theory with $U(1)$ gauge symmetry, Lorentz space-time symmetry,
the discrete charge, parity, and combined CP symmetries,
can be constructed as
\bea
{\cal L}_{eff} &=& {\cal L}_2 + {\cal L}_4 +  \cdots + {\cal L}_{qd}\,,\label{efflag}\\
{\cal L}_2  &=&  {v^2\over 2} (D U)^{\dagger}\cdot(D U)\,,\\
{\cal L}_4 &=& - {1\over 4 e^2} F_{\mn}F^{\mn}
+ d_1 [(D U)^{\dagger}\cdot(D U)]^2 \,,\\
\cdots\,,\nnb\\
{\cal L}_{qd} &=&  i \delta(0) \left \{
e_1 {(D U)^{\dagger}\cdot(D U) \over m_1^2}
+e_2 {[(D U)^{\dagger}\cdot(D U)]^2 \over m_1^4}
+ \cdots\right  \} \,,
\eea
where the ${\cal L}_2 $ contains relevant operator, the
${\cal L}_4$ contains marginal operators, and the irrelevant
operators are represented by the dots. The $m_1$ is the mass of vector bosons,
and $m_1= e v$. The term $[(D U)^{\dagger}\cdot(D U)]$
in the ${\cal L}_2$ reflects the effects of $U(1)$ symmetry breaking (the vector
bosons obtained their masses).
And the effective coupling $d_1$ in the ${\cal L}_4$ is one of the anomalous couplings
if judging from the viewpoint of the renormalizable $U(1)$ gauge
theory. The operators in ${\cal L}_2$ and ${\cal L}_4$ belong to the operators
up to $O(p^4)$ in the usual momentum counting rule.
The effective Lagrangian (\ref{efflag}) is invariant under
the following $U(1)$ gauge transformation
\bea
U\rightarrow \exp(-i \alpha) U\,,\,\,A_{\mu}\rightarrow A_{\mu} + \partial_{\mu} \alpha\,.
\eea

Matching the full theory with the effective theory
by integrating out the massive scalar field
$\rho$, we can fix the effective
couplings at the matching scale $\mu=m_0$.
Since we only consider the integrating-out at tree
level at the matching scale, the equation of motion of the heavy Higgs
boson $\rho$, which expresses the $\rho$ in the low energy DOFs, is sufficient
for this purpose, which reads
\bea
\label{eom}
\rho = {v \over m_0^2} (D U)^{\dagger}\cdot(D U) + \cdots\,,
\eea
where $m_0$ is the mass of Higgs boson.
The omitted terms contain at least four covariant partials.
The effective couplings of operators at the matching scale $m_0$
are fixed and listed below
\bea
\label{init}
d_1(m_0) = {v^2 \over 2 m_0^2}={1\over \lambda}\,\,,\,\,
e_1(m_0) = - {m_1^2 \over m_0^2}\,\,,\,\,
e_2(m_0) = - {m_1^4 \over 2 m_0^4}\,\,,\,\,
\cdots\,.
\eea
The $d_1(m_0)$ vanishes in the decoupling limit $m_0^2\rightarrow\infty$,
and it is to reflect the remnants of the high energy dynamics.

It is remarkable that terms in
the ${\cal L}_{qd}$ must be kept in the effective
Lagrangian, and they are nonvanishing and can be derived
directly from the Lagrangian density of
the non-linear Higgs model given in Eqn. (\ref{uglag}).
The spurious argument to drop these quaritic divergence terms
from the effective Lagrangian, by regarding that these terms are infinite
and have no physical signigicance, can not hold,
since, as shown in the U-gauge, they do play a necessary
part to cancel the quartic divergences produced by the longitudinal
part of vector propagators.
Such a term is also crucial for the
elimination of quartic divergences coming from the loop
corrections in the effective theory, as will be shown below.
In effect we can drop all quartic divergences from the
beginning, but to include them and find the calcellation mechanism
seems more consistent in a quantum field theory.

To consider the logarithmic contributions of the
low energy DOFs to the effective couplings of the
${\cal L}_{eff}$ given in Eqn. (\ref{efflag}),
we will use the RGE method to sum them.
Compared with the procedure given in \cite{std}, where
the decoupling limit is taken before the direct one-loop level
matching step is made to extract the logarithmic contributions,
our procedure is more consistent in the
framework of effective theory and is rather simpler.

To derive the RGEs of the effective theory,
we conduct our calculation in the background field method (BFM) \cite{bfm}.
In this method, the vector boson fields are split into background and quantum parts,
which are represented as ${\overline A}$ and ${\widehat A}$,
respectively. The Goldstone field is also split into two parts,
\bea
U={\overline U} {\widehat U}\,\,,\,\,{\widehat U}=\exp\left( \frac{i \xi}{v}\right)\,\,,
\eea
but the background Goldstone field ${\overline U}$
can be absorbed in the redefinition of the background vector fields
${\overline A}$ by invoking the Stueckelberg transformation \cite{stuck}
\bea
{\overline A_{\mu}^s} \rightarrow {\overline {A_{\mu}}} + i \partial_{\mu} {\overline U}\,,
\eea
and will not appear in the effective Lagrangian any more.
After integrating-out the quantum fields, by using the inverse
Stueckelberg transformation
we can restore the effective Lagrangian in its low energy dynamic variables of
the ${\overline A}$ and ${\overline U}$.

As one of the advantages of the BFM, we have the freedom to choose
different gauges for the background part and the quantum part.
To formualted the quadratic terms of quantum fields into the
standard form given in Eq. (\ref{stdf}), the gauge fixing term of the
quantum fields can be chosen as
\bea
{\cal L}_{GF} = -{1\over 2 e^2} (\partial\cdot \widehat A + f_{A\xi} \xi)^2\,,
\eea
where $f_{A\xi}= e^2 v$.

Up to one loop level, the partition functional of the effective Lagrangian
in the background gauge can be expressed as
\bea
\label{effpar}
{\cal Z} = \exp\left (i {\cal S}^{ren}[{\overline A^s}]\right ) = \exp\left (i {\cal S}_{tree}[{\overline A^s}] + i \delta {\cal S}_{tree}[{\overline A^s}] + i {\cal S}_{1-loop}[{\overline A^s}] + \cdots\right )\,\nnb\\
=\exp\left (i {\cal S}_{tree}[{\overline A^s}] + i \delta{\cal S}_{tree}[{\overline A^s}]\right ) \int {\cal D}{\widehat A_{\mu}} {\cal D}\xi \exp\left (i {\cal S}[{\widehat A},\xi;{\overline A^s}]\right )\,,
\eea
where the tree effective Lagrangian ${\cal L}_{tree}$ is in the following form
\bea
{\cal L}_{tree}&=&  -{1\over 4 e^2} F_{\mn}F^{\mn}
+ {v^2\over 2} {\overline A^s}\cdot{\overline A^s}
+ d_1 ({\overline A^s}\cdot{\overline A^s})^2
+ \cdots\,
\nnb\\
&&+ i \delta(0) \left (
  e_1 { {\overline A^s}\cdot{\overline A^s}\over m_1^2}
 + e_2 {({\overline A^s}\cdot{\overline A^s})^2\over m_1^4}
+ \cdots \right  ) \,,
\eea
And the counter terms of the effective Lagrangian $\delta {\cal L}_{tree}$ are defined as
\bea
\delta {\cal L}_{tree} &=&  -\delta Z_{e^2} {1\over 4 e^2} F_{\mn}F^{\mn}
+ \delta Z_{v^2} {v^2\over 2} {\overline A^s}\cdot{\overline A^s}
+ \delta Z_{d_1} d_1 ({\overline A^s}\cdot{\overline A^s})^2
+ \cdots\nnb\\
&&+ i \delta(0) \left (
  \delta e_1 { {\overline A^s}\cdot{\overline A^s}\over m_1^2}
+ \delta e_2 {({\overline A^s}\cdot{\overline A^s})^2\over m_1^4}
+ \cdots \right  ) \,,
\label{ctt}
\eea
where the renormalization constant of ${\overline A^s}$ is set to $1$.

We would like to comment on the partition functional
given in Eq. (\ref{effpar}). According to the matching procedure prescribed
by H. Georgi \cite{georgi}, it is the Green functions of the full and effective
theories that should be matched. Equivalently, we can
match the effective irreducible vertice functional.
In fact we can indeed conduct the procedure
of integrating out and matching in the BFM from the full theory at tree level,
which yields the same result as given in Eq. (\ref{init}).

Up to one-loop corrections, only quadratic terms in
${\cal S}[{\widehat A},\xi;{\overline A^s}]$ are important, which can
be organized as
\bea
\label{stdf}
{\cal L}_{quad}&=&
{1\over 2 } {\widehat A_{\mu}} \Box^{\mu\nu}_{A\,A} {\widehat A_{\nu}}
+ {1\over 2} \xi  \Box_{\xi\,\xi} {\partial}^2 \xi}
+ {1\over 2} {\widehat A_{\mu}} {X^{\mu}_{A\,\xi}} \xi
+ {1\over 2} \xi {X^{\nu}_{\xi\,A} {\widehat A_{\nu}}\,,\\
\Box^{\mu\nu}_{A\, A} &=& ({\partial}^2 + m_1^2) g^{\mu\nu} - \sigma^{\mu\nu}\,\,,\\
\Box_{\xi\,\xi} &=& ({\partial}^2 + m_1^2) -X^{\alpha} \partial_{\alpha} - X^{\alpha\beta} \partial_{\alpha} \partial_{\beta}\,\,,\\
\sigma^{\mu\nu}_{A\,A}    &=& - 4 e^2 d_1 DA^{\mu\nu}\,\,,\\
 X^{\alpha}        &=& -{4 d_1 \over v^2} \partial_{\beta} DA^{\mu\nu}\,\,,\\
 X^{\alpha\beta}   &=& -{4 d_1 \over v^2} DA^{\alpha\beta}\,\,,\\
 X^{\mu}_{A\,\xi}  &=&  X^{\mu\alpha}_{A\,\xi} \partial_{\alpha} + X^{\mu}_{A\,\xi; 03}\,\,,\\
 X^{\nu}_{\xi\,A}  &=&  X^{\nu\alpha}_{\xi\,A} \partial_{\alpha} + X^{\nu}_{\xi\,A; 03}\,\,,\\
 X^{\mu\alpha}_{A\,\xi}  &=& -{4 e d_1 \over v} DA^{\mu\alpha}\,\,,\\
 X^{\nu\alpha}_{\xi\,A}  &=&  {4 e d_1 \over v} DA^{\nu\alpha}\,\,,\\
 X^{\mu}_{A\,\xi; 03}    &=& -{1\over 2} \partial_{\alpha} X^{\mu\alpha}_{A\,\xi}\,\,,\\
 X^{\nu}_{\xi\,A; 03}    &=& -{1\over 2} \partial_{\alpha} X^{\nu\alpha}_{\xi\,A}\,\,,\\
 DA^{\mu\nu}  &=& {\overline A^s}\cdot{\overline A^s} g^{\mu \nu} + 2 {\overline A^s}^{\mu} {\overline A^s}^{\nu}\,\,.
\eea
Here we have performed the diagonalization transformation
${\widehat A} \rightarrow {\widehat A}/e$ to make the quadratic terms
of quantum vector fields in their standard forms.
Now in oder to extract operators only up to the relevant and the marginal
($O(p^4)$) operators, we introduce an auxiliary dimension counting rule, which
reads
\bea
[{\overline A}]_a=[\partial]_a=1\,,\,\,[v]_a=0\,.
\eea
This auxiliary counting rule gives
\bea
[X^{\alpha}]_a = [X^{\mu}_{A\,\xi; 03}]_a = [X^{\nu}_{\xi\,A; 03}]_a = 3\,\,,\,\,
[\sigma^{\mu\nu}]_a = [X^{\alpha\beta}]_a = [X^{\mu\alpha}_{A\,\xi}]_a =[X^{\nu\alpha}_{\xi\,A}]_p=2\,.
\eea
Since there is no terms with dimension 1 in the standard form, then terms with dimension 3
will not contribute to the one-loop effective Lagrangian up to $O(p^4)$.
Therefore below we will neglect them in our calculations.
We would like to point out that this auxiliary dimension counting rule
by itself is to extract terms with two and four external fields,
if speaking in the language of Feynman diagrams.

The quadratic terms can be directly calculated in the functional integral.
Then after integrating out all quantum fields, the ${\cal L}_{1-loop}$ reads
\bea
\int_x {\cal L}_{1-loop}=i {1\over 2} \left [Tr\ln\Box_{A\,A} +Tr\ln\Box_{\xi\,\xi} +Tr\ln\left (1-X_{\xi\,A}^{\mu} \Box^{-1}_{A\,A;\mu\nu} X_{A\,\xi}^{\nu} \Box^{-1}_{\xi\,\xi} \right  )\right]\,.
\label{logtr}
\eea
This form is quite compact, but we need to expand it and extract those
terms up to $O(p^4)$.

With the divergence and auxiliary dimension counting rules,
it is direct to evaluate the logarithm and trace given in Eqn. (\ref{logtr})
up to the relevant and marginal ($O(p^4)$) order by both the Feynman rules
method in the background field method and the covariant heat kernel method.
Considering that the Schwinger proper time and heat kernel method is
explicit covariant and are simpler than the Feynman rule method in the non-Abelian cases,
we demonstrate the covariant heat kernel method (In the $U(1)$ case, these
two methods are different in the representative spaces, and the former
method has no much advantage over the latter one).
Here the divergence counting rule is just the ordinary Dyson power
counting rule (in the coordinate space). In fact, these two rules can
help us to extract the divergences up to any a specified order.

Below we conduct our calculation in the Euclidean space.
In the Schwinger proper time and the heat kernel method, the propagators can be expressed as
\bea
\langle x|\Box^{-1}_{A\,A;\mu\nu}|y\rangle=\int_0^{\infty} \frac{d \lambda}{(4 \pi \lambda)^{d\over 2}} \exp\left(- m_1^2 \lambda \right) \exp\left( - {z^2\over 4 \lambda}\right) H^{\mu\nu}_{A\,A}(x,y;\lambda)\,\,,\\
\langle x|\Box^{-1}_{\xi\,\xi}|y \rangle=\int_0^{\infty} \frac{d \lambda}{(4 \pi \lambda)^{d\over 2}} \exp\left(- m_1^2 \lambda \right) \exp\left( - {z^2\over 4 \lambda}\right) H_{\xi\,\xi}(x,y;\lambda)\,\,,
\eea
where $z=y-x$. The $H_{A\,A}$ and $H_{\xi\,\xi}$
can be expanded with reference to
$\lambda$, and the corresponding coefficients are
called Silly-De Witt coefficients \cite{ball, avra}.

The $Tr\ln\Box_{A\,A}$ can only contribute to quadratic and
logarithm divergences, and can be evaluated in the standard
minimal subtraction scheme, which yields
\bea
Tr\ln\Box_{A\,A} &=& - {1 \over (4 \pi)^2} {2\over \epsilon} \int_x  \left \{
- m_1^2 tr(\sigma_{A\,A}^{\mu\nu} g_{\mu\nu})
+{1\over 2} tr(\sigma_{A\,A}^{\mu\nu} \sigma_{A\,A,\nu\mu}) \right \}\,.
\eea

The $Tr\ln\Box_{\xi\,\xi}$ can be expressed as
\bea
Tr\ln\Box_{\xi\,\xi} = Tr\ln ( \partial^2 + m_1^2 )
+ Tr\ln \left ( 1 - X^{\alpha\beta} \partial_{\alpha} \partial_{\beta} (\partial^2 + m_1^2)^{-1} \right)\,,
\eea
where the first term can be neglected since it contains no external field and
only contribute to the unobservable vacuum.
The second term contributes to the quartic divergences and can be expanded as
\bea
Tr\ln( 1 - X^{\alpha\beta}\partial_{\alpha} \partial_{\beta}(\partial^2 + m_1^2)^{-1})&=&
- Tr X^{\alpha\beta}\partial_{\alpha} \partial_{\beta} (\partial^2 + m_1^2)^{-1}\nnb\\
&&+ {1\over 2} Tr \left[ X^{\alpha\beta}\partial_{\alpha} \partial_{\beta} (\partial^2 + m_1^2)^{-1}\right]^2
+ \cdots\,,
\label{trace}
\eea
where the first term in the rhs of Eqn (\ref{trace}) is proportional
to ${ {\overline A^s}\cdot{\overline A^s}/m_1^2}$, and the second
one is proportional to ${[{\overline A^s}\cdot{\overline A^s}]^2/m_1^4}$.
We can always adjust the $\delta e_1$ and $\delta e_2$ (which are finite) in
the counter terms $\delta{\cal L}_{tree}$ given in (\ref{ctt}) to
guarantee that $e_1^{ren}$ and $e_2^{ren}$ vanish. So in practical
calculation quartic divergences can be simply thrown away.

The $Tr\ln\left (1-X_{\xi\,A}^{\mu} \Box^{-1}_{A\,A;\mu\nu} X_{A\,\xi}^{\nu} \Box^{-1}_{\xi\,\xi} \right )$
can be expanded as
\bea
\label{mxtr}
 Tr\ln\left (1-X_{\xi\,A}^{\mu} \Box^{-1}_{A\,A;\mu\nu} X_{A\,\xi}^{\nu} \Box^{-1}_{\xi\,\xi} \right  )&=&
-Tr X_{\xi\,A}^{\mu} \Box^{-1}_{A\,A;\mu\nu} X_{A\,\xi}^{\nu} \Box^{-1}_{\xi\,\xi}\nnb\\
&&+{1\over 2} Tr (X_{\xi\,A}^{\mu} \Box^{-1}_{A\,A;\mu\nu} X_{A\,\xi}^{\nu} \Box^{-1}_{\xi\,\xi})^2
+\cdots\,\,.
\eea
The first term in the rhs of Eqn. (\ref{mxtr}) is quadratically divergent, the
auxiliary dimension counting rule indicates that it contributes to the operators in $O(p^4)$,
$O(p^6)$, and so on.
While the second term is logarithm divergent, and the auxliary dimension counting rule
indicates it contributes to the operators in $O(p^8)$, $O(p^{10})$, and so on.
So here we see one of the characteristics
of the non-linear interaction---that low dimension operators can induce
infinite number of divergences of higher dimension operators. Such a fact is
also quite explicit if the calculation is conducted in the
unitary gauge. However, since we only consider the
effective theory up to the $O(p^4)$ order, we can drop those higher dimension
divergences. The justification for this practice will be explained below.

Now we use the Schwinger proper time and the heat kernel method to calculate
the trace of the first term of the rhs of Eqn. (\ref{mxtr}). In the
coordinate space, the trace can be expressed as
\bea
II&=&Tr X_{\xi\,A}^{\mu} \Box^{-1}_{A\,A;\mu\nu} X_{A\,\xi}^{\nu} \Box^{-1}_{\xi\,\xi}\nnb\\
&=&\langle x | X_{\xi\,A}^{\mu} |x\rangle \langle x | \Box^{-1}_{A\,A;\mu\nu} |y\rangle \langle y| X_{A\,\xi}^{\nu} |y \rangle \langle y|\Box^{-1}_{\xi\,\xi}|x\rangle\,.
\eea
Keeping only terms up to $O(p^4)$, we can write $II$ as
\bea
II= - e^2 \int_{z,\lambda_1,\lambda_2} {g_{\mn} \over (4 \pi)^d} { X_{\xi\,A}^{\mu\alpha} X_{A\,\xi}^{\nu\beta}\over (\lambda_1 \lambda_2)^{d\over 2}} {z_{\alpha} z_{\beta}  \over 4 \lambda_1 \lambda_2}
 \exp\left[ - m_1^2 (\lambda_1 + \lambda_2) \right]
\exp\left[ - {z^2 \over 4}  ({1\over \lambda_1} + {1\over \lambda_2} ) \right]\,.
\eea
After integrating all variables, we get
\bea
II = -{1\over 2}  X_{\xi\,A}^{\mu\alpha} X_{A\,\xi}^{\nu\beta} g_{\mu\nu} g_{\alpha\beta}\,.
\eea

Combining all contributions, we get ${\cal L}_{1-loop}$
\bea
{\cal L}_{1-loop} = {1\over (4 \pi)^2} {2\over \epsilon}
\left[ 12 d_1 v^2 e^4 {\overline A^s}.{\overline A^s} \right]
+\cdots\,,
\eea
and from it, we can construct the counter terms and extract renormalization constants.
With the renormalization constants, we can get the RGEs of the effective theory, which read
\bea
\label{rge}
\frac{d v^2}{d t} &=& - 48 d_1 \alpha_e^2 v^2\,,
\eea
where $\alpha_e=e^2/(4 \pi)$. It's remarkable
that the gauge coupling $e$ and the effective coupling $d_1$ do not run.
The reason is that there is no logarithmic divergence of the gauge kinetic term,
since the $U(1)$ guage contains no self-interaction.
While for $d_1$, the reason is that
the logarithmic divergences of the $d_1$ term from the $Tr\ln\Box_{A\,A}$
and the $Tr\ln\left (1-X_{\xi\,A}^{\mu} \Box^{-1}_{A\,A;\mu\nu} X_{A\,\xi}^{\nu} \Box^{-1}_{\xi\,\xi} \right )$
cancel exactly.

The RGE given in Eq. (\ref{rge}) can be solved exactly,
and the solutions read
\bea
v^2(t) &=& v_{UV}^2 \exp\left(-48 \alpha_e^2 d_1 t\right)\,,
\eea
It's apparent that the radiative corrections of low energy DOF tend to increase the value of $v^2$ when the
sliding scale of the effective theory run from its ultraviolet (UV) cutoff
down to its infrared cutoff, the magnitude of the change depends
on the values of $e$, $v_{UV}$, and, $d_{1,UV}$.

The RGE method of an effective theory is an improvement on the perturbation
calculation, and will make it simple, at least at the one-loop level, to
match the effective and full theories.
Without using the RGE method, the matching directly at the
one-loop level is quite complicated.

We have shown in the effective $U(1)$ gauge theory as how to
conduct the renormalization at $O(p^4)$ order. As we have emphasized, that
the quartic divergent terms must be added in the effective Lagragian in order
to eliminate the quartic divergences generated by the loops of low energy DOF.
And those two counting rules---the divergence and auxiliary dimension counting rules---
are quite helpful extract the terms up to a specified order.

In principle, the procedure we shown above can be extended
to discuss any specified higher order corrections. As stated in the
reference \cite{wein}, if we take into account all permitted
operators, it is still possible to do the renormalization of an
effective theory as we do in the renormalizable one without
any proximity to truncate the infinite operator series. However,
according to the predictions of the Wilson renormalization
method \cite{wilson} and the surface theorem found by
Polchinski \cite{pol}, the higher the dimension of the operator,
the smaller is the strength of the couplings at low energy scale.
So for a specified accuracy, when the
low energy scale is smaller enough than the UV cutoff scale, we
can safely neglect the contributions
of those higher dimension operators (even there are divergences originated
from low energy DOFs), and only take into account
the effects of those finite important operators. That's why we can
truncate the infinite divergences tower to the $O(p^4)$ order.

We have checked the above methods to extract divergent
structures by computing directly in the Feynman-Dyson method with the usual
Feynman rules and Feynman diagrams,
which yields the same results.
In the example of the U(1) theory, we
have conducted our computation in the Feynman and
't hooft gauge, and in fact, it is also quite simple to calculte in the
U-gauge. However, in the non-Abelian cases,
considering that fact that the U-gauge will
bring into double poles \cite{wood, sonoda} and
we will stick to use the Feynman and 't hooft gauge.
The conceptions and method can easily extended to the
non-Abeliant cases. Using the background field method,
Stueckelberg transformation, those two counting rules,
and, the method to evaluate the logarithm and trace, the
RGEs of non-linear effective non-Abelian theories with spontaneous
symmetry breaking can also be derived \cite{qsyan}.

{\bf Acknowledgments:}
One of the author, Q. S. Yan, would like to thank
Professor C. D. L\"u and Dr. X. J. Bi for helpful discussion.
Especial thanks are given for Professor Y. P. Kuang,
who has helped to find the right auxiliary dimension
counting rule and made a good comment on the matching
procedure, and for Professor Q. Wang, who has helped
in improving the presentation of this paper. The work of Q. S. Yan is
supported by the Chinese Postdoctoral Science Foundation
and the CAS K. C. Wong Postdoctoral Research Award Foundation.
The work of D. S. Du is supported by the
National Natural Science Foundation of China.


\begin{thebibliography}{10}
\bibitem{georgi}
	H. Georgi, Annu. Rev. Nucl. Part. Sci. {\bf 43} (1993) 209.
\bibitem{pich}
	A. Pich, hep-ph/9806303.
\bibitem{buras}
	G. Buchalla, A. J. Buras, and M. E. Lauthenbache, Rev. Mod. Phys. 68 (1996) 1125.
\bibitem{wilson}
	K. G. Wilson and J. Kogut, Phys. Repts. 12 C, (1974) 75.
\bibitem{gasser}
	J. Gasser, and H. Leutwyler, Nucl. Phys. B 250 (1985) 465;
	Ann. Phys. 158 (1984) 142.	
\bibitem{appliq}
	A. C. Longhitano, Phys. Rev. D 22 (1980) 1166; Nucl. Phys. B 188 (1981) 118;
	T. Appelquist, and G. H. Wu, Phys. Rev. D 48 (1993) 3235.
\bibitem{higgs}
	P. W. Higgs, Phys. Rev. 145 (1966) 1156.
\bibitem{nish}
	K.Nishijima, and T. Watanabe; Prog. Theor. Physics. Vol 49 (1973) 341.
\bibitem{wood}
	K. A. Woodhouse, Nuovo Cimento A 23 (1973) 459;
	C. Grosse-Knetter, and R. K\"ogerler, Phys. Rev. D 48 (1993) 2865.
\bibitem{std}
	M. J. Herrero, and E. R. Morales, Nucl. Phys. B 418 (1994) 431;
	S. Dittmaier and C. Grosse-Knetter, hep-ph/9505266 and hep-ph/9501285.
\bibitem{bfm}
	B.S. DeWitt, Phys. Rev. 162(1967)1195, 1239.
\bibitem{stuck}
	E. C. G. Stueckelberg., Helv. Phys. Acta 11 (1938) 299; 30 (1956) 209;
	T. Kunimasa and T. Goto, Prog. Theor. Phys. 37 (1967) 524.
\bibitem{ball}
	R. D. Ball, Phys. Rep. 182 (1989) 1.
\bibitem{avra}
	I. G. Avramidi, {\it Lecture Notes in Physics: N.s. M. Monogrph; 64}
	{\bf Heat Kernel and Quantum Gravity}, Springer, 2000.
\bibitem{wein}
	S. Weinberg, The Quantum Theory of Fields, Vol. I, Published by Cambridge University
	Press, Chapter 12.
\bibitem{pol}
	J. Polchinski, Nucl. Phys. B 231 (1984) 269.
\bibitem{sonoda}
	H. Sonoda, hep-th/0108217.
\bibitem{qsyan}
	Q. S. Yan, and D. S. Du, in preparation.
\end{thebibliography}
\end{document}